\documentclass[preprint,12pt]{elsarticle}
\journal{Composite Structures}
\usepackage{graphics,epsfig}
\usepackage{graphicx}
\usepackage{float}
\usepackage{amssymb,amstext,amsmath}
\usepackage{mathtools}
\usepackage{booktabs}
\usepackage{natbib}
\usepackage{physics}
\setcitestyle{round,aysep={,},yysep={,},citesep={;}}
\usepackage{listings}
\newcommand{\operator}[1]{\mathop{\vphantom{\sum}\mathchoice
{\vcenter{\hbox{\huge $#1$}}}
{\vcenter{\hbox{\Large $#1$}}}{#1}{#1}}\displaylimits}
\newcommand{\opA}{\operator{\mathrm{A}}}

\newcommand{\balpha}{\boldsymbol{\alpha}}
\newcommand{\bvarepsilon}{\boldsymbol{\varepsilon}}
\newcommand{\bchi}{\boldsymbol{\chi}}
\newcommand{\bxi}{\boldsymbol{\xi}}
\newcommand{\bnu}{\boldsymbol{\nu}}
\newcommand{\bpsi}{\boldsymbol{\psi}}
\newcommand{\llbracket}{[\![}
\newcommand{\rrbracket}{]\!]}
\newcommand{\llangle}{\langle \! \langle}
\newcommand{\rrangle}{\rangle \! \rangle}

\begin{document}
\begin{frontmatter}
\title{Asymptotically exact theory of fiber-reinforced composite beams}
\author{K. C. Le$^{a,b}$\footnote{Email: lekhanhchau@tdtu.edu.vn}, T. M. Tran$^{a,b}$}
\address{$^a$Materials Mechanics Research Group, Ton Duc Thang University, Ho Chi Minh City, Vietnam\\
$^b$Faculty of Civil Engineering, Ton Duc Thang University, Ho Chi Minh City, Vietnam}
\begin{abstract} 
An asymptotic analysis of the energy functional of a fiber-reinforced composite beam with a periodic microstructure in its cross section is performed. From this analysis the asymptotically exact energy as well as the 1-D beam theory of first order is derived. The effective stiffnesses of the beam are calculated in the most general case from the numerical solution of the cell and homogenized cross-sectional problems.
\end{abstract}


\end{frontmatter}

\section{Introduction}

Fibre reinforced composite (FRC) beams are widely used in civil, mechanical, and aerospace engineering due to their low weight, high strength, and good damping properties \citep{hodges1990review,chandra1999damping,librescu2005thin}. The exact treatment of FRC beams within the 3D elasticity is only possible in a few exceptional cases due to their complicated microstructure (see, e.g., \citep{rodriguez2001closed,sendeckyj2016mechanics} and the references therein). For this reason, different approaches have been developed depending on the type of beams. If FRC beams are thick, no exact one-dimensional theory can be constructed, so only the numerical methods or approximate semi-analytical methods applied to three-dimensional elasticity theory make sense. However, if the FRC beams are thin, the reduction from the three-dimensional to the one-dimensional theory is possible. This dimension reduction can be made asymptotically exact in the limit when the thickness-to-length ratio of the beam goes to zero. The rigorous derivation of the asymptotically exact one-dimensional beam theory based on the variational-asymptotic method (VAM) developed by \citet{berdichevsky1979variational} was first performed in \citep{berdichevsky1981on}. His asymptotic analysis shows that the static (or dynamic) three-dimensional problem of the beam can be split into two problems: (i) the two-dimensional cross-sectional problem and (ii) the one-dimensional variational problem whose energy functional should be found from the solution of the cross-sectional problem. The latter has been solved both for anisotropic homogeneous beams and for inhomogeneous beams with the constant Poisson's ratio \citep{berdichevsky1981on}. In addition to these findings, \citet{berdichevsky1981on} has shown that the average energy as well as the extension and bending stiffnesses of FRC beams with piecewise constant Poisson's ratio must be larger than those of FRC beams with constant Poisson's ratio (see also \citep{muskhelishvili2013some}). However, to our knowledge, the question of how the corrections in energy and stiffnesses depend on the difference in Poisson's ratio of fibers and matrix of FRC beams remains still an issue. It should be noted that VAM has been further developed in connection with the numerical analysis of cross-sectional problems for geometrically nonlinear composite beams by Hodges, Yu and colleagues in a series of papers \citep{hodges1992on,yu2002on,yu2005a,yu2012a}.  Note also that VAM has been used, among others, to derive the 2D theory of homogeneous piezoelectric shells \citep{Le86a}, the 2D theory of purely elastic anisotropic and inhomogeneous shells \citep{berdichevsky2009variational}, the 2D theory of purely elastic sandwich plates and shells  \citep{berdichevsky2010asymptotic,berdichevsky2010nonlinear}, the theory of smart beams  \citep{roy2007asymptotically}, the theory of low and high frequency vibrations of laminate composite shells \citep{lee2009adynamic,lee2009bdynamic}, and more recently, the theory of smart sandwich and functionally graded shells \citep{le2016asymptotically,le2017an}.  

For FRC beams that have the periodic microstructure in the cross section, an additional small parameter appears in the cross-sectional problems: The ratio between the length of the periodic cell and the characteristic size of the cross section. In this case the finite element code VABS developed in the above mentioned papers \citep{hodges1992on,yu2002on,yu2005a,yu2012a} cannot be applied to the cross-sectional problem because it requires a large computational capacity. The presence of this small parameter allows however an additional asymptotic analysis of the cross-sectional problems to simplify them. By solving the cell problems imposed with the periodic boundary conditions according to the homogenization technique \citep{braides2002gamma,milton2003theory}, one finds the effective coefficients in the homogenized cross-sectional problems, which can then be solved analytically or numerically (cf. also \citep{liu2017two}). The aim of this paper is to derive and solve the cell and homogenized cross-sectional problems for unidirectional FRC beams whose cross section has the periodic microstructure. For simplicity, we will assume that both matrix and fibers are elastically isotropic but have different Poisson's ratio. The solution of the cell problems found with the finite element method is used to calculate the asymptotically exact energy and the extension and bending stiffnesses in the 1-D theory of FRC beams. Thus, we determine the dependence of the latter quantities on the shape and volume fraction of the fibers and on the difference in Poisson's ratio of fibers and matrix that solves the above mentioned issue.

The paper is organized as follows. After this short introduction the variational formulation of the problem is given in Section 2. Sections 3 and 4 are devoted to the multi-scaled asymptotic analysis of the energy functional of FRC beams leading to the cross-sectional and cell problems. In Section 5 the cell problems are solved by the finite element method. Section 6 provides the solutions of the homogenized cross-sectional problems. Section 7 present one-dimensional theory of FRC beams. Finally, Section 8 concludes the paper.

\section{Variational formulation for FRC beams}
\begin{figure}[htb]
	\centering
	\includegraphics[width=5cm]{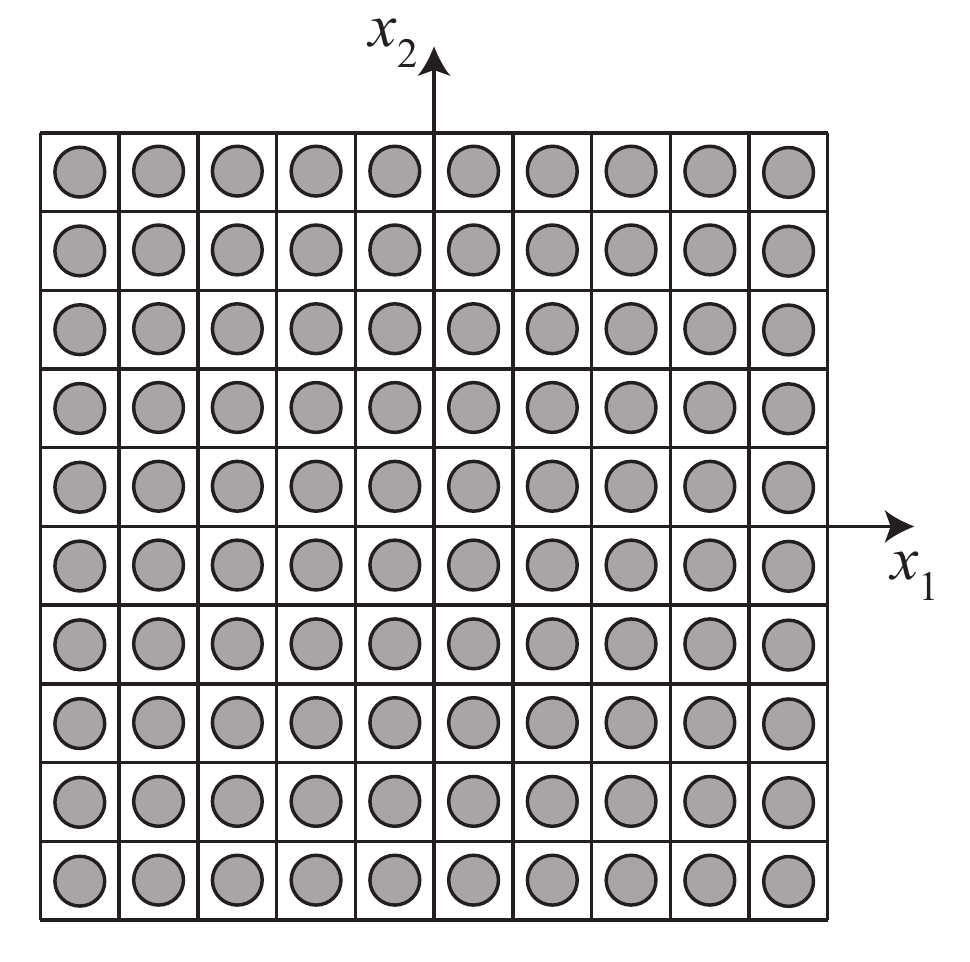}
	\caption{A cross section of a FRC beam}
	\label{fig:1}
\end{figure}
Consider a FRC beam that occupies the domain $\mathcal{B}=\mathcal{A}\times (0,L)$ of the 3-D euclidean space in its undeformed state. Let $x_3\equiv x$ be the coordinate along the beam axis. The cross section of the beam in the $(x_1,x_2)$-plane, $\mathcal{A}$, consist of two separated 2-D sub-domains $\mathcal{A}^m$ and $\mathcal{A}^f$ such that the matrix occupies the domain $\mathcal{A}^m\times (0,L)$, while the uni-directional fibers occupy the domain $\mathcal{A}^f\times (0,L)$. We choose the origin of the $(x_1,x_2)$-coordinates so that it matches the centroid of $\mathcal{A}$. We assume that the fibers are periodically situated in the matrix and the bond between the fibers and the matrix is perfect (see Fig.~\ref{fig:1} representing the cross section of the beam where $\mathcal{A}^f$ is the set of gray circles). We first consider the beam in equilibrium, whose deformation is completely determined by the displacement  field ${\bf u}(\mathbf{x})$. Let, for simplicity, the edge $x=0$ of the beam be clamped, while at the other edge $x=L$ of the beam the traction $\mathbf{t}(x_1,x_2)$ be specified. Gibbs' variational principle \citep{berdichevsky2009variational} states that the true displacement $\check{\mathbf{u}}(\mathbf{x})$ minimizes the energy functional
\begin{equation}
\label{2.2}
I[\mathbf{u}(\mathbf{x})]=\int_{\mathcal{B}}W(\mathbf{x},\bvarepsilon)\dd[3]{x}-\int_{\mathcal{A}} \mathbf{t}(x_1,x_2)\vdot \mathbf{u}(x_1,x_2,L) \dd[2]{x}
\end{equation} 
among all admissible displacements $\mathbf{u}(\mathbf{x})$ satisfying the boundary conditions
\begin{displaymath}
\mathbf{u}(x_1,x_2,0)=0.
\end{displaymath}
Here, $\dd[3]{x}=\dd x_1\dd x_2\dd x$ is the volume element, $\dd[2]{x}=\dd x_1\dd x_2$ the area element, and the dot denotes the scalar product. Function $W(\mathbf{x},\bvarepsilon)$, called stored energy density, reads 
\begin{equation*}
W(\mathbf{x},\bvarepsilon)=\frac{1}{2}\lambda(\mathbf{x}) (\tr \bvarepsilon)^2+\mu(\mathbf{x}) \bvarepsilon \mathbf{:}\bvarepsilon,
\end{equation*}
where $\boldsymbol{\varepsilon} $ is the strain tensor
\begin{equation*}
\bvarepsilon =\frac{1}{2}(\grad \mathbf{u}+(\grad \mathbf{u})^T).
\end{equation*}

The problem is to replace the three-dimensional energy functional by an approximate one-dimensional energy functional for a thin FRC beam, whose functions depend only on the longitudinal co-ordinate $x$. The possibility of reduction of the three- to the one-dimensional functional is related to the smallness of two parameters: (i) the ratio between the thickness $h$ of the cross section and the length $L$ of the beam, and (ii) the ratio between the size $c$ of the periodic cell and $h$. By using the variational-asymptotic method, the one-dimensional energy functional will be constructed below in which terms of the order $h/L$ and $\epsilon=c/h$ are neglected as compared with unity (the first-order or ``classical'' approximation). Formally, this corresponds to the limits $h\to 0$ and $\epsilon
\to 0$.

In order to perform the variational-asymptotic analysis, it is convenient to use the index notation, with Greek indices running from 1 to 2 and denoting the component of vectors and tensors in the $(x_1,x_2)$-plane. Summation over repeated Greek indices is understood. Index 3 of the coordinate $x_3$, the displacement $u_3$, and the traction $t_3$ is dropped for short. To fix the domain of the transverse co-ordinates in the passage to the limit $h\to 0$, we introduce the dimensionless co-ordinates
\begin{equation*}
y_\alpha =\frac{x_\alpha }{h}, \quad y_\alpha \in \bar{\mathcal{A}},
\end{equation*}
and transform the energy functional of the beam to
\begin{equation}\label{3.4}
I=\int_{0}^{L}\int_{\bar{\mathcal{A}}}h^2W(y_\alpha ,\bvarepsilon ) \dd[2]{y} \dd{x} -\int_{\bar{\mathcal{A}}} h^2\mathbf{t}(y_\alpha)\vdot \mathbf{u}(y_\alpha,L) \dd[2]{y}.
\end{equation}
The transverse coordinates $y_\alpha $ play the role of the ``fast'' variables as opposed to the slow variable $x$. Regarding $\vb{u}$ as function of $y_\alpha$ and $x$, we separate the fast variables  from the slow one. Now $h$ enters the action functional explicitly through the components of the strain tensor $\bvarepsilon $ 
\begin{equation}\label{3.5} 
\varepsilon _{\alpha \beta }=\frac{1}{h}u_{(\alpha ;\beta )}, \quad
2\varepsilon _{\alpha 3}= 
\frac{1}{h}u_{;\alpha }+u_{\alpha,x},
\quad
\varepsilon_{33}=u_{,x}.
\end{equation}
Here and below, the semicolon preceding Greek indices denotes the derivatives with respect to $y_\alpha$, while the parentheses surrounding a pair of indices the symmetrization operation. 

Besides $y_\alpha$ there are also much faster variables
\begin{displaymath}
z_\alpha =\frac{x_\alpha}{c}=\frac{y_\alpha}{\epsilon}
\end{displaymath} 
associated with the periodicity of the elastic moduli of this composite materials in the transverse directions leading to the fast oscillation of the stress and strain fields in these directions. In order to characterize these faster changes of the stress and strain fields in the transverse directions we will assume that the displacement field $\vb{u}$ is a composite function of $\vb{y}$ and $x$ such that
\begin{displaymath}
\vb{u}=\vb{u}(\vb{y},\vb{y}/\epsilon,x)=\vb{u}(\vb{y},\vb{z},x),
\end{displaymath} 
where $\vb{y},\vb{z}$ are two-dimensional vectors with the components $y_\alpha $, $z_\alpha $, respectively, and where $\vb{u}(\vb{y},\vb{z},x)$ is a doubly periodic function in $\vb{z}$ with the period 1. This is a typical multi-scale Ansatz to composite structures. The asymptotic analysis must therefore be performed twice, first in the limit $h\to 0$ to realize the dimension reduction, and then in the limit $\epsilon\to 0$ to homogenize the cross-sectional problems and solve them. 

\section{Dimension reduction}

Before starting the asymptotic analysis of the energy functional in the limit $h\to 0$ let us transform the stored energy density to another form more convenient for this purpose \citep{le1999vibrations}. We note that, among terms of $W(\vb{z}, \bvarepsilon )$, the derivatives $w_{(\alpha ;\beta )}/h$ in $\varepsilon _{\alpha \beta}$ and $w_{;\alpha }/h$ in $\varepsilon _{\alpha 3}$ are the main ones in the asymptotic sense. Therefore it is convenient to single out the components $\varepsilon _{\alpha \beta}$ and $\varepsilon _{\alpha 3}$ in the stored energy density. We represent the latter as the sum of three quadratic forms $W_\parallel $, $W_\angle$, and $W_\perp $ according to
\begin{equation*}
W_\parallel =\min_{\varepsilon _{\alpha \beta},
\varepsilon _{\alpha 3}}W, 
\quad
W_\angle =\min_{\varepsilon _{\alpha \beta}} (W-W_\parallel ), 
\quad
W_\perp =W-W_\parallel -W_\angle .
\end{equation*}
The ``longitudinal'' energy density $W_\parallel $ depends only on $\varepsilon _{33}$ and coincides with $W$ when the stresses $\sigma _{\alpha \beta }$ and $\sigma _{\alpha 3}$ vanish; the ``shear'' energy density $W_\angle $ depends only on $\varepsilon _{\alpha 3}$; the remaining part $W_\perp $ is called the ``transverse'' energy density. From the definitions of $W_\parallel $, $W_\angle $ and
$W_\perp $ one easily finds out that
\begin{gather*}
W_\parallel =\frac{1}{2}E(\varepsilon _{33})^2, \quad 
E=\frac{\mu (3\lambda +2\mu )}{\lambda +\mu },
\\ 
W_\angle =2\mu \varepsilon 
_{\alpha 3}\varepsilon _{\alpha 3}, 
\\
W_\perp =\frac{1}{2}\lambda (\varepsilon _{\alpha \alpha}
+2\nu \varepsilon _{33})^2+ \mu (\varepsilon _{\alpha \beta }
+\nu \delta _{\alpha \beta }\varepsilon _{33}) (\varepsilon _{\alpha \beta }
+\nu \delta _{\alpha \beta }\varepsilon _{33}),
\end{gather*}
where $E$ is the Young modulus and $\nu =\lambda /2(\lambda +\mu )$ the Poisson ratio. Note that $E$, $\nu$ as well as Lame's  constants $\lambda$, $\mu$ are doubly periodic functions of the fast variables $\vb{z}$. Note also the following identities
\begin{align*}
W_\angle &= \sigma_{\alpha 3}\varepsilon_{\alpha 3}, \notag \\
W_\perp &= \frac{1}{2} (\sigma_{\alpha \beta} \varepsilon_{\alpha \beta}+\sigma_{33}\varepsilon_{33}-E(\varepsilon _{33})^2). 
\end{align*}

We could start the variational-asymptotic analysis in the limit $h\to 0$ with the determination of the set $\mathcal{N}$ according to its general scheme \citep{le1999vibrations}. As a result, it would turn out that, at the first step, the function ${\bf u}$ does not depend on the transverse co-ordinates $\vb{y}$ (and $\vb{z}$): ${\bf u}={\bf v}(x)$; at the second step the function ${\bf u}^\star $ becomes a linear function of $\vb{y}$ and involves one more degree of freedom $\varphi (x)$ representing the twist angle; and at the next step ${\bf u}^{\star \star }$ is completely determined through ${\bf u}$ and $\varphi$. Thus, the set $\mathcal{N}$ according to the variational-asymptotic scheme consists of functions ${\bf v}(x)$ and $\varphi(x)$. We will pass over these long, but otherwise standard, deliberations and make a change of unknown functions according to 
\begin{equation}\label{5.2}
\begin{split}
u_\alpha (\vb{y},\vb{z},x)=v_\alpha (x)-he_{\alpha \beta}\varphi (x)y_\beta+hw _\alpha (\vb{y},\vb{z},x),
\\
u(\vb{y},\vb{z},x)=v(x)-h v_{\alpha ,x} (x) y_\alpha +hw(\vb{y},\vb{z},x),
\end{split} 
\end{equation}
where $e_{\alpha \beta }$ are the two-dimensional permutation symbols ($e_{11}=e_{22}=0,e_{12}=-e_{21}=1$). By redefining $\vb{v}(x)$ and $\varphi (x)$ if necessary we can impose on functions $w_\alpha $ and $w$ the following constraints
\begin{equation}\label{5.3}
\begin{split}
\langle w_\alpha \rangle =0,\quad  e_{\alpha \beta 
} \langle w_{\alpha ;\beta }\rangle =0, \\
\quad \langle w \rangle =0, \quad  \langle . \rangle \equiv \int_{\bar{\mathcal{A}}} . \dd[2]{y}.
\end{split}
\end{equation}
According to these constraints $v_\alpha (x)$ and $v(x)$ describe the mean displacements of the beam, while $\varphi (x)$ corresponds to the mean rotation of its cross section. Equations \eqref{5.2} and \eqref{5.3} set up a one-to-one correspondence between $u_\alpha ,u$ and the set of functions $v_\alpha ,v, \varphi, w_\alpha ,w$ and determine the change in the unknown functions $\{ u_\alpha ,u\} \to \{ v_\alpha ,v, \varphi, w_\alpha ,w \}$.

Based on the Saint-Venant principle for elastic beams, we may assume that the domain occupied by the beam consists of the inner domain and two boundary layers near the edges of the beam with width of the order $h$ where the stress and strain states are essentially three-dimensional. Then functional \eqref{3.4} can be decomposed into the sum of two functionals, an inner one for which an iteration process will be applied, and a boundary layer functional. When searching for $ w_\alpha $ and $w$ the boundary layer functional can be neglected in the first-order approximation. Therefore, the dimension reduction problem reduces to finding the minimizer $\check{w}_\alpha$ and $\check{w}$ of the inner functional that, in the limit $h\to 0$, can be identified with the functional \eqref{3.4} without the last term.

We now fix $v_\alpha ,v,\varphi $ and seek $w_\alpha ,w$. Substituting \eqref{5.2} into the action functional \eqref{3.4} with the last term being removed, we will keep in it the asymptotically principal terms containing $w_\alpha ,w $ and neglect all other terms. The estimations based on Eqs.~\eqref{3.5} and \eqref{5.2} lead to the asymptotic formulas
\begin{equation}
\varepsilon _{\alpha \beta }=w_{(\alpha ;\beta )},\quad 2\varepsilon _{\alpha 3}=
w_{;\alpha }-he_{\alpha \beta}\Omega y_\beta,\quad \varepsilon _{33}=\gamma+h\Omega_\alpha y_\alpha,
\label{5.6}
\end{equation}
where $\gamma$, $\Omega_\alpha$, and $\Omega$ are the measures of elongation, bending, and twist defined by
\begin{equation*}
\gamma=v_{,x},\quad \Omega_\alpha =-v_{\alpha,xx},\quad \Omega=\varphi_{,x}.
\end{equation*}
According to equations \eqref{5.6} the partial derivatives of $w_\alpha , w $ with respect to 
$x$ do not enter the energy functional. As $x$ becomes the formal parameter, we may drop the integral over $x$ and reduce the determination of $w_\alpha , w$ to the uncoupled minimization problems for every fixed $x$ of the functionals
\begin{multline}\label{5.8}
I_\perp [w_\alpha ]=h^2 \langle \frac{1}{2}\lambda(\vb{z}) [w_{\alpha ;\alpha } 
+2\nu(\vb{z}) (\gamma +h\Omega _\sigma y_\sigma )]^2 
\\
+\mu(\vb{z}) [w_{(\alpha ;\beta )} +\nu(\vb{z}) \delta _{\alpha \beta }(\gamma +h\Omega _\gamma y_\gamma )][w_{(\alpha ;\beta )} +\nu(\vb{z}) \delta _{\alpha \beta }(\gamma +h\Omega _\delta y_\delta )]\rangle ,
\end{multline}
\begin{equation}\label{5.9} 
I_\angle [w] =h^2 \langle \frac{1}{2}\mu(\vb{z}) (w_{;\alpha }-h\Omega e_{\alpha 
\beta}y_\beta)(w_{;\alpha }-h\Omega e_{\alpha \gamma}y_\gamma)\rangle .
\end{equation}
The minima are searched among all admissible functions $w_\alpha , w $ satisfying the constraints \eqref{5.3}. Note that the decoupling of problems \eqref{5.8} and \eqref{5.9} holds true in the most general case of anisotropy \citep{berdichevsky2009variational}. This can be seen from the asymptotically main terms containing the unknown functions $w_\alpha$ and $w$ in the transverse and shear strain energy densities: $w_{;\alpha}$ does not enter $W_\perp$ while $w_{\alpha ;\beta }$ does not enter $W_\angle $. Functionals \eqref{5.8} and \eqref{5.9} represent the transverse and shear strain energies, integrated over the cross section of the beam. They are positive definite and convex, so the existence of their minimizers $\check{w}_\alpha ,\check{w}$ is guaranteed. We shall see in the next Sections that the minimum of \eqref{5.8} is equal to zero if $\nu$ is equal for both matrix and fiber, while that of \eqref{5.9} is equal to $1/2C\Omega ^2$, with $C$ the torsional rigidity.

\section{Homogenization}

Consider now the other limit $\epsilon\to 0$. In this limit $\vb{y}$ plays the role of the ``slow'' variable, while $\vb{z}=\vb{y}/\epsilon$ becomes the fast variable. We start with the cross-sectional problem of minimizing functional \eqref{5.9} among $w(\vb{y},\vb{z},x)$ satisfying the constraint \eqref{5.3}$_3$, where $\mu (\vb{z})$, expressed in terms of the fast variable $\vb{z}=\vb{y}/\epsilon $, is a doubly periodic function with period 1. Since $x$ is fixed in this cross-sectional problem, we shall drop this formal variable of $w$ in this Section. Following the homogenization technique based also on the variational-asymptotic method \citep{berdichevsky2009variational}, we look for the minimizer in the form
\begin{equation}\label{eq:comp2}
w(\vb{y},\vb{z})=\psi(\vb{y})+\epsilon \chi (\vb{z}),
\end{equation}
where $\chi (\vb{z})$ is a doubly periodic function with period 1. Note that $\chi $ may depend on the slow variable $\vb{y}$, but this dependency is suppressed for short. In addition to the constraint \eqref{5.3}$_3$ we may impose the following constraint on $\chi (\vb{z})$
\begin{equation}
\label{eq:comp5}
\llangle \chi \rrangle \equiv \int_{\mathcal{C}}\chi (\vb{z})\dd[2]{z}=0,
\end{equation}
where $\mathcal{C}=(0,1)\times (0,1)$ is the unit periodic cell. In this case $\psi (\vb{y})$ can be interpreted as the average value of $w(\vb{y},\vb{z})$ over the cell. Although the second term in \eqref{eq:comp2} is small compared to $\psi(\vb{y})$ and goes to zero in the limit $\epsilon \to 0$, it contribution to the shear strains as the gradient of $w(\vb{y},\vb{z})$ has the same order as the gradient of $\psi(\vb{y})$ as seen from the asymptotic formula
\begin{equation}
\label{eq:comp7}
w_{;\alpha }=\psi_{;\alpha }+\chi_{|\alpha }, \quad \alpha =1,2.
\end{equation}
Here, the vertical bar followed by an index $\alpha $ denotes the partial derivative with respect to the corresponding fast variable $z_\alpha $.

Inserting \eqref{eq:comp7} into the energy functional \eqref{5.9}, we get
\begin{equation*}
I_\angle [\psi(\vb{y}),\chi (\vb{y}/\epsilon)]=h^2 \langle \frac{1}{2}\mu (\vb{y}/\epsilon)(\chi_{|\alpha}+\xi_\alpha )(\chi_{|\alpha}+\xi_\alpha )\rangle ,
\end{equation*}
with $\xi_\alpha=\psi_{;\alpha}-h\Omega e_{\alpha \beta}y_\beta$ being the function of the ``slow'' variable $\vb{y}$. We replace the double integral $\langle .\rangle$ by the sum of double integrals over the cells. Then
\begin{equation}
\label{eq:comp9}
I_\angle [\psi(\vb{y}),\chi (\vb{y}/\epsilon)]=h^2 \sum_{n=1}^N \int_{\mathcal{C}_n} \frac{1}{2}\mu (\vb{y}/\epsilon)(\chi_{|\alpha}+\xi_\alpha )(\chi_{|\alpha}+\xi_\alpha )\dd[2]{y},
\end{equation}
where $N$ is the total number of cells and $\mathcal{C}_n$ is the cell $((i-1)\epsilon,i \epsilon)\times ((j-1)\epsilon,j \epsilon)$. We minimize this functional in two steps: (i) Fix $\psi(\vb{y})$ and minimize the functional among doubly periodic $\chi(\vb{y}/\epsilon)$, (ii) Minimize the obtained functional among $\psi(\vb{y})$. Because $\chi(\vb{y}/\epsilon)$ is doubly periodic with respect to $\vb{y}$, we can minimize each integral in the sum independently. Besides, as $\xi_\alpha(\vb{y})$ change slowly in one cell, we may regard them in each cell integral as constants equal to their value in the middle of the cell. It is convenient to change the variable $\vb{y}$ in the cell integrals to $\vb{z}=\vb{y}/\epsilon -(i-1,j-1)$. Then the minimization of cell integrals reduces to the following cell problem: Minimize the functional
\begin{equation}
\label{eq:comp10}
\llangle \frac{1}{2}\mu(\vb{z})(\chi_{|\alpha}+\xi_\alpha )(\chi_{|\alpha}+\xi_\alpha )\rrangle
\end{equation}
among doubly periodic functions $\chi(\vb{z})$ satisfying the constraint \eqref{eq:comp5}, where, as before, $\llangle .\rrangle$ is the double integral in $\vb{z}$ over the unit cell. We denote the obtained minimum by $\bar{W}_\angle(\xi_\alpha )$ and call it the average shear energy density. Then, replacing the sum in \eqref{eq:comp9} by the integral in the limit $\epsilon \to 0$, we arrive at the following homogenized cross-sectional problem: Minimize the average functional
\begin{equation}\label{eq:comp11}
\bar{I}_\angle[\psi (\vb{y})]=h^2 \langle \bar{W}_\angle (\psi_{;\alpha}-h\Omega e_{\alpha 
\beta}y_\beta)\rangle 
\end{equation}
among $\psi (\vb{y})$ satisfying the condition 
\begin{equation*}
\langle \psi \rangle =0.
\end{equation*}

The standard calculus of variations shows that the minimizer of \eqref{eq:comp10} satisfies the equation
\begin{equation}
\label{eq:comp12a}
[\mu(\vb{z})(\chi_{|\alpha }+\xi_\alpha )]_{|\alpha }=0 
\end{equation}
in the unit cell, where $\bxi$ can be regarded as the constant vector in the cell (since it does not depend on the fast variable $\vb{z}$). This equation is subjected to the periodic boundary condition and the constraint \eqref{eq:comp5}. Besides, as the shear modulus suffers a jump at the boundary $\partial_i$ between the fiber and the matrix, the continuity of the traction
\begin{equation}
\label{eq:comp12b}
\llbracket \mu(\vb{z})(\chi_{|\alpha }+\xi_\alpha )\rrbracket \nu _\alpha =0
\end{equation}
should be fulfilled at this internal boundary $\partial_i$, where $\bnu$ is the unit normal vector outward to the fiber and $\llbracket . \rrbracket $ denotes the jump.

After finding the minimizer $\check{\chi }(\vb{z})$ as a solution to the boundary-value problem \eqref{eq:comp12a}-\eqref{eq:comp12b}, we substitute it back into functional \eqref{eq:comp10} to calculate the average shear energy density
\begin{align}
\bar{W}_\angle(\xi_{\alpha})&=\llangle \frac{1}{2}\mu(\vb{z})(\check{\chi}_{|\alpha}+\xi_{\alpha})(\check{\chi}_{|\alpha}+\xi_{\alpha})\rrangle \notag
\\
&=\llangle \frac{1}{2}\mu(\vb{z})(\check{\chi}_{|\alpha}+\xi_{\alpha})\check{\chi}_{|\alpha}\rrangle 
+\llangle \frac{1}{2}\mu(\vb{z})(\check{\chi}_{|\alpha}+\xi_{\alpha})\xi_{\alpha}\rrangle . \label{eq:comp12c}
\end{align}
As the minimizer $\check{\chi}$ satisfies \eqref{eq:comp12a}-\eqref{eq:comp12b}, the first integral must vanish. Using the constancy of $\xi_{\alpha}$, we reduce the second integral to
\begin{equation*}
\bar{W}_\angle(\xi_{\alpha})=\frac{1}{2}\xi_{\alpha} \llangle \mu(\vb{z})(\check{\chi}_{|\alpha}+\xi_{\alpha})\rrangle .
\end{equation*}
The integrand is the shear stress $\sigma_{3\alpha}$, so the integral gives the average shear stress $\bar{\sigma}_{3\alpha}$. On the other side, due to the constraint \eqref{eq:comp5},
\begin{equation*}
\llangle \check{\chi}_{|\alpha}+\xi_{\alpha} \rrangle = \xi_{\alpha} ,
\end{equation*}
so $\xi_{\alpha}$ is the average engineering shear strain $\bar{\varepsilon}_{3\alpha }$. We define the effective elastic shear moduli $\mu ^*_{\alpha \beta}$ by the linear equation
\begin{equation}
\label{eq:comp12f}
\bar{\sigma}_{3\alpha}=\llangle \mu(\vb{z})(\check{\chi}_{|\alpha}+\xi_{\alpha})\rrangle  =\mu ^*_{\alpha \beta}\xi_{\beta} .
\end{equation}
Note that the homogenized material must not necessarily be isotropic even if the components of the composite are. So, in general $\mu ^*_{\alpha \beta}$ is a tensor of second order. With this we get the average shear energy density in terms of $\xi_{\alpha}=\psi_{;\alpha}-h\Omega e_{\alpha \beta}y_\beta$
\begin{equation}
\label{eq:comp12g}
\bar{W}_\angle(\xi_{\alpha})=\frac{1}{2}\mu ^*_{\alpha \beta} \xi_{\alpha}\xi_{\beta}.
\end{equation}
We want to show that the tensor $\mu ^*_{\alpha \beta}$ in \eqref{eq:comp12f} is  symmetric. Indeed, since $\bar{W}_\angle(\xi_\alpha)$ in \eqref{eq:comp12g} is a quadratic form, we can replace $\mu ^*_{\alpha \beta}$ there by the symmetric tensor $\mu ^*_{(\alpha \beta)}=\frac{1}{2}(\mu ^*_{\alpha \beta}+\mu ^*_{\beta \alpha })$. If we substitute \eqref{eq:comp12g} into \eqref{eq:comp12c} and differentiate it with respect to $\xi_{\alpha}$, we get
\begin{displaymath}
\mu ^*_{(\alpha \beta)}\xi _\beta =\llangle \mu(\vb{z})(\check{\chi}_{|\alpha}+\xi_\alpha)\rrangle =\mu ^*_{\alpha \beta}\xi_\beta ,
\end{displaymath}
which proves the symmetry of $\mu ^*_{\alpha \beta}$.

We turn now to the cross-sectional problem \eqref{5.8}. Before doing the asymptotic analysis for it in the limit $\epsilon\to 0$ let us prove the following remarkable property: The minimum of functional \eqref{5.8} is zero  if the Poisson ratio $\nu$ is an equal constant for both fiber and matrix \citep{berdichevsky1981on}. Indeed, let us choose the minimizer in the form
\begin{equation}
\label{eq:comp13}
\check{w}_\alpha =-\nu \gamma \delta_{\alpha \beta}y_\beta -\frac{1}{2}\nu a_{\alpha \beta \gamma}( y_\beta y_\gamma -\langle y_\beta y_\gamma\rangle /|\bar{\mathcal{A}}|) ,
\end{equation}
where $|\bar{\mathcal{A}}|$ is the area of $\bar{\mathcal{A}}$ and
\begin{displaymath}
a_{\alpha \beta \gamma}=h(\delta_{\alpha \beta}\Omega_\gamma +\delta_{\alpha \gamma}\Omega_\beta-\delta_{\beta \gamma}\Omega_\alpha ).
\end{displaymath}
Note that the tensor $a_{\alpha \beta \gamma}$ is symmetric with respect to the last two indices, and
\begin{displaymath}
\check{w}_{\alpha ;\beta }=-\nu \gamma \delta_{\alpha \beta }-a_{\alpha \beta \gamma}y_\gamma.
\end{displaymath}
Due to our choice of the origin, $\langle y_\alpha \rangle =0$. Therefore the chosen field \eqref{eq:comp13} satisfies the constraints \eqref{5.3}$_{1,2}$. It is easy to check that the transverse energy evaluated at $\check{w}_\alpha $ vanishes identically. Since functional \eqref{5.8} is non-negative definite, its minimum is obviously zero in this case.

The asymptotic analysis of problem \eqref{5.8} for changeable $\nu$ is quite similar to that of problem \eqref{5.9}. In this case the minimizer is sought in the form
\begin{equation*}
w_\alpha (\vb{y},\vb{z})=\psi_\alpha (\vb{y})+\epsilon \chi _\alpha (\vb{z}),
\end{equation*}
where $\chi_\alpha$ satisfy the constraints
\begin{equation}\label{eq:comp15}
\llangle \chi_\alpha \rrangle =0.
\end{equation}
To determine $\chi_\alpha $ we need to solve the following cell problem: Minimize the functional
\begin{multline}
\label{eq:comp16}
\llangle \frac{1}{2}\lambda(\vb{z}) (\chi_{\alpha |\alpha } +\bar{\varepsilon}_{\alpha \alpha}
+2\nu(\vb{z}) \xi )^2 
\\
+\mu(\vb{z}) (\chi_{(\alpha |\beta )} +\bar{\varepsilon}_{\alpha \beta}+\nu(\vb{z}) \delta _{\alpha \beta }\xi )(\chi_{(\alpha |\beta )} +\bar{\varepsilon}_{\alpha \beta}
+\nu(\vb{z}) \delta _{\alpha \beta }\xi) \rrangle
\end{multline}
among doubly periodic functions $\chi_\alpha (\vb{z})$ satisfying the constraint \eqref{eq:comp15}, where
\begin{displaymath}
\bar{\varepsilon}_{\alpha \beta}=\psi_{(\alpha |\beta)}, \quad \xi =\gamma +h\Omega _\alpha y_\alpha. 
\end{displaymath}
Note that this problem is quite similar to the problem of determining the effective thermal expansion of composite material with periodic microstructure, where $\xi $ plays the role of the temperature increase \citep{Sigmund1997}. Let the minimum of functional \eqref{eq:comp16} be $\bar{W}_\perp(\bar{\varepsilon}_{\alpha \beta},\xi)$. Then the determination of $\psi_\alpha$ reduces to minimizing the following average functional
\begin{equation}\label{eq:comp17}
\bar{I}_\perp[\psi_\alpha (\vb{y})]=h^2 \langle \bar{W}_\perp (\psi_{(\alpha;\beta)},\gamma +h\Omega _\alpha y_\alpha)\rangle 
\end{equation}
among $\psi_\alpha (\vb{y})$ satisfying the constraints 
\begin{equation}\label{eq:comp17a}
\langle \psi_\alpha \rangle =0,\quad e_{\alpha \beta 
} \langle \psi_{\alpha ;\beta }\rangle =0.
\end{equation}

The minimizer of functional \eqref{eq:comp16} satisfy the equilibrium equations
\begin{equation*}
[\lambda (\vb{z})(\chi_{\beta|\beta}+\bar{\varepsilon}_{\beta \beta}+\xi)]_{|\alpha}+[2\mu(\vb{z})(\chi_{(\alpha|\beta)}+\bar{\varepsilon}_{\alpha \beta})]_{|\beta }=0 
\end{equation*}
in the unit cell, where $\bar{\varepsilon}_{\alpha \beta }$ and $\xi$ can be regarded as the constant tensor and scalar in the cell (since they do not depend on the fast variable $\vb{z}$). These equations are subjected to the periodic boundary conditions and the constraints \eqref{eq:comp15}. Besides, as the elastic moduli suffers jumps at the boundary $\partial_i$ between the fiber and the matrix, the continuity of the traction
\begin{equation*}
\llbracket \lambda (\vb{z})(\chi_{\beta|\beta}+\bar{\varepsilon}_{\beta \beta}+\xi)\nu_\alpha +2\mu(\vb{z})(\chi_{(\alpha|\beta)}+\bar{\varepsilon}_{\alpha \beta} )\nu _\beta \rrbracket =0
\end{equation*}
should be fulfilled there. Since the functional \eqref{eq:comp16} is quadratic, its minimum must be a quadratic form of $\bar{\varepsilon}_{\alpha \beta}$ and $\xi$
\begin{equation}
\label{eq:comp21}
\bar{W}_\perp(\bar{\varepsilon}_{\alpha \beta},\xi)=\frac{1}{2}C^*_{\alpha \beta \gamma \delta}\bar{\varepsilon}_{\alpha \beta}\bar{\varepsilon}_{\gamma \delta}+D^*_{\alpha \beta }\bar{\varepsilon}_{\alpha \beta}\xi+\frac{1}{2}F^* \xi ^2.
\end{equation}
Using the same arguments as in the previous case, we can prove the following symmetry properties of the effective moduli
\begin{gather*}
C^*_{\alpha \beta \gamma \delta} =  C^*_{\beta \alpha \gamma \delta}=C^*_{\alpha \beta \delta \gamma }=C^*_{\gamma \delta \alpha \beta}, \\
D^*_{\alpha \beta} = D^*_{\beta \alpha }.
\end{gather*}
We also introduce the effective tensor of Poisson's ratios by
\begin{equation*}
\nu^*_{\alpha \beta}=C^{*(-1)}_{\alpha \beta \gamma \delta}D_{\gamma \nu},
\end{equation*}
where $C^{*(-1)}_{\alpha \beta \gamma \delta}$ is the tensor of elastic compliances defined by
\begin{equation*}
C^{*(-1)}_{\alpha \beta \gamma \delta}C_{\gamma \delta \zeta \eta}=\delta_{\alpha (\zeta }\delta_{\beta \eta)}.
\end{equation*}
It is easy to see that $\nu_{\alpha \beta}$ is symmetric.

\section{Numerical solution of the cell problems}

It is convenient to rewrite functional \eqref{eq:comp16} by changing the sign of $\chi _\alpha(\vb{y})$ 
\begin{equation}\label{eq:numeric1}
\frac{1}{2}\llangle C_{\alpha \beta \gamma \delta}(\vb{z})(-\chi_{(\alpha |\beta)}+\bar{\varepsilon}_{\alpha \beta}+\alpha _{\alpha \beta }(\vb{z})\xi)(-\chi_{(\gamma |\delta)}+\bar{\varepsilon}_{\gamma \delta}+\alpha _{\gamma \delta }(\vb{z})\xi)\rrangle , 
\end{equation}
where 
\begin{equation*}
C_{\alpha \beta \gamma \delta}(\vb{z})=\lambda (\vb{z})\delta_{\alpha \beta}\delta_{\gamma \delta}+\mu(\vb{z})(\delta_{\alpha \gamma}\delta_{\beta \gamma}+\delta_{\alpha \delta }\delta_{\beta \gamma}), \quad \alpha _{\alpha \beta }(\vb{z})=\nu(\vb{z})\delta_{\alpha \beta}.
\end{equation*}
The minimizer of functional \eqref{eq:numeric1} satisfies the variational equation
\begin{equation}
\label{eq:numeric2}
\llangle C_{\alpha \beta \gamma \delta}(\vb{z})\chi_{(\gamma |\delta )}\delta \chi_{\alpha |\beta} \rrangle =\llangle C_{\alpha \beta \gamma \delta}(\vb{z})(\bar{\varepsilon}_{\gamma \delta}+\alpha _{\gamma \delta }(\vb{z})\xi )\delta \chi_{\alpha |\beta} \rrangle 
\end{equation}
for all doubly periodic functions $\delta \chi _\alpha$. Eq.~\eqref{eq:numeric2} will be solved by the finite element method \citep{hughes2012the,andreasen2014how}. For this purpose it is convenient to change from tensor notation to matrix notation \citep{hughes2012the}, in which Eq.~\eqref{eq:numeric2} becomes
\begin{equation}
\label{eq:numeric2a}
\llangle \bvarepsilon^T(\bpsi)\vb{C}(\vb{z})\bvarepsilon(\bchi) \rrangle =\llangle \bvarepsilon^T(\bpsi)\vb{C}(\vb{z})(\bar{\bvarepsilon}+\balpha(\vb{z})\xi) \rrangle ,
\end{equation}
where $\bpsi=\delta \bchi$, $\bvarepsilon(\bchi)$ is the strain vector given by
\begin{displaymath}
\bvarepsilon(\bchi)=(\chi_{1|1},\chi_{2|2},\chi_{1|2}+\chi_{2|1})^T,
\end{displaymath}
while
\begin{equation}\label{eq:numeric4}
\vb{C}(\vb{z})=\lambda (\vb{z})\begin{bmatrix}
   1   &  1  &  0  \\
   1  &   1  &  0  \\
   0  &  0  &  0
\end{bmatrix} +\mu(\vb{z})\begin{bmatrix}
   2   &  0  &  0  \\
   0  &   2  &  0  \\
   0  &  0  &  1
\end{bmatrix} ,\quad \balpha(\vb{z})=\nu (\vb{z})\begin{bmatrix}
    1 \\  1  \\  0  
\end{bmatrix}.
\end{equation}
The discretization of this equation based on the bilinear isoparametric elements is standard. The periodic boundary conditions are imposed by identifying the nodes on opposite sides of the unit cell. This is implemented by using the matrix edofMat for a full, regular grid to index into a periodic version of the grid \citep{andreasen2014how}. The global stiffness matrix is
\begin{equation*}
\vb{K}=\opA_{e=1}^N (\vb{k}_e), \quad \vb{k}_e=\int_{A_e} \vb{B}^T_e\vb{C}_e\vb{B}_e\dd[2]{z} ,
\end{equation*}
where $\opA$ is the assembly operator taken over the total number $N$ of finite elements. The matrix $\vb{B}_e$ is the element strain-displacement matrix, $A_e$ is the domain of element $e$, and $\vb{C}_e$ is the constitutive matrix for the element. The indicator matrix is introduced that specifies whether the element is in $\mathcal{A}^m$ ($n_e=1$) or in $\mathcal{A}^f$ ($n_e=2$). The piecewise constant $\vb{C}(\vb{z})$ from \eqref{eq:numeric4} takes value $\vb{C}_e$ in the element in accordance with this indicator matrix.

The discretization of the right hand side of \eqref{eq:numeric2a} yields the loads $\vb{f}^{(i)}$ 
\begin{equation*}
\vb{f}^{(i)}=\opA_{e=1}^N (\vb{f}^{(i)}_e), \quad \vb{f}^{(i)}_e=\int_{A_e}\vb{B}^T_e\vb{C}_e \bar{\bvarepsilon}^{(i)} \dd[2]{z}, \quad i=1,2,3
\end{equation*}
which correspond to the average macroscopic strains
\begin{equation}
\label{eq:numeric6}
\bar{\bvarepsilon}^{(1)}=(1,0,0)^T,\quad \bar{\bvarepsilon}^{(2)}=(0,1,0)^T, \quad \bar{\bvarepsilon}^{(3)}=(0,0,1)^T,
\end{equation}
and 
\begin{equation*}
\vb{f}^{(4)}=\opA_{e=1}^N (\vb{f}^{(4)}_e), \quad \vb{f}^{(4)}_e=\int_{A_e}\vb{B}^T_e\vb{C}_e \balpha_e \dd[2]{z}, 
\end{equation*}
which correspond to $\xi=1$, with $\balpha_e$ being the constitutive vector from \eqref{eq:numeric4} that takes values in the element in accordance with the indicator matrix. The displacement fields are computed by solving the corresponding linear equations with four load-cases
\begin{equation*}
\vb{K}\bchi^{(i)}=\vb{f}^{(i)},  \quad i=1,2,3,4.
\end{equation*}

When the displacements are obtained, the elements of the effective matrix $\bf{C}^*$ are found as:
\begin{equation*}
C^*_{ij}=\sum_{e=1}^N \int_{A_e}(\bchi_e^{0(i)}-\bchi_e^{(i)})^T \vb{k}_e(\bchi_e^{0(j)}-\bchi_e^{(j)})\dd[2]{z}, \quad i,j=1,2,3,
\end{equation*}
where $\bchi_e^{0(i)}$ are the three displacement fields corresponding to the unit strains from \eqref{eq:numeric6}, and $\bchi_e^{(i)}$ contains three columns corresponding to the three displacement fields resulting from globally enforcing the unit strains \eqref{eq:numeric6}. The indices in parentheses refer to the column number. The components of the effective moduli $D^*_i$ ($D^*_{\alpha \beta }$) due to $\xi $ are computed according to
\begin{equation*}
D^*_{i}=\sum_{e=1}^N \int_{A_e}(\balpha_e-\vb{B}_e \bchi_e^{(4)})^T \vb{C}_e(\bar{\bvarepsilon}^i-\vb{B}_e \bchi_e^{(i)})\dd[2]{z}, \quad i=1,2,3.
\end{equation*}
Having $D^*_i$ we can also compute the effective ``Poisson'' ratios $\nu^*_i$ ($\nu^*_{\alpha \beta}$) as follows
\begin{equation}
\label{eq:numeric9}
\nu^*_{i}=(C^*_{ij})^{-1}D^*_{j}.
\end{equation}
Finally, the effective coefficient $F^*$ is given by
\begin{equation*}
F^*=\sum_{e=1}^N \int_{A_e}(\balpha_e-\vb{B}_e \bchi_e^{(4)})^T \vb{C}_e(\balpha_e-\vb{B}_e \bchi_e^{(4)})\dd[2]{z}.
\end{equation*}
The Matlab-code homogenizecs.m to solve the cell problem and compute the effective moduli which is a modification of the code homogenize.m written by \citet{andreasen2014how} is presented in the Appendix.

\begin{figure}[htb]
	\centering
	\includegraphics[width=7cm]{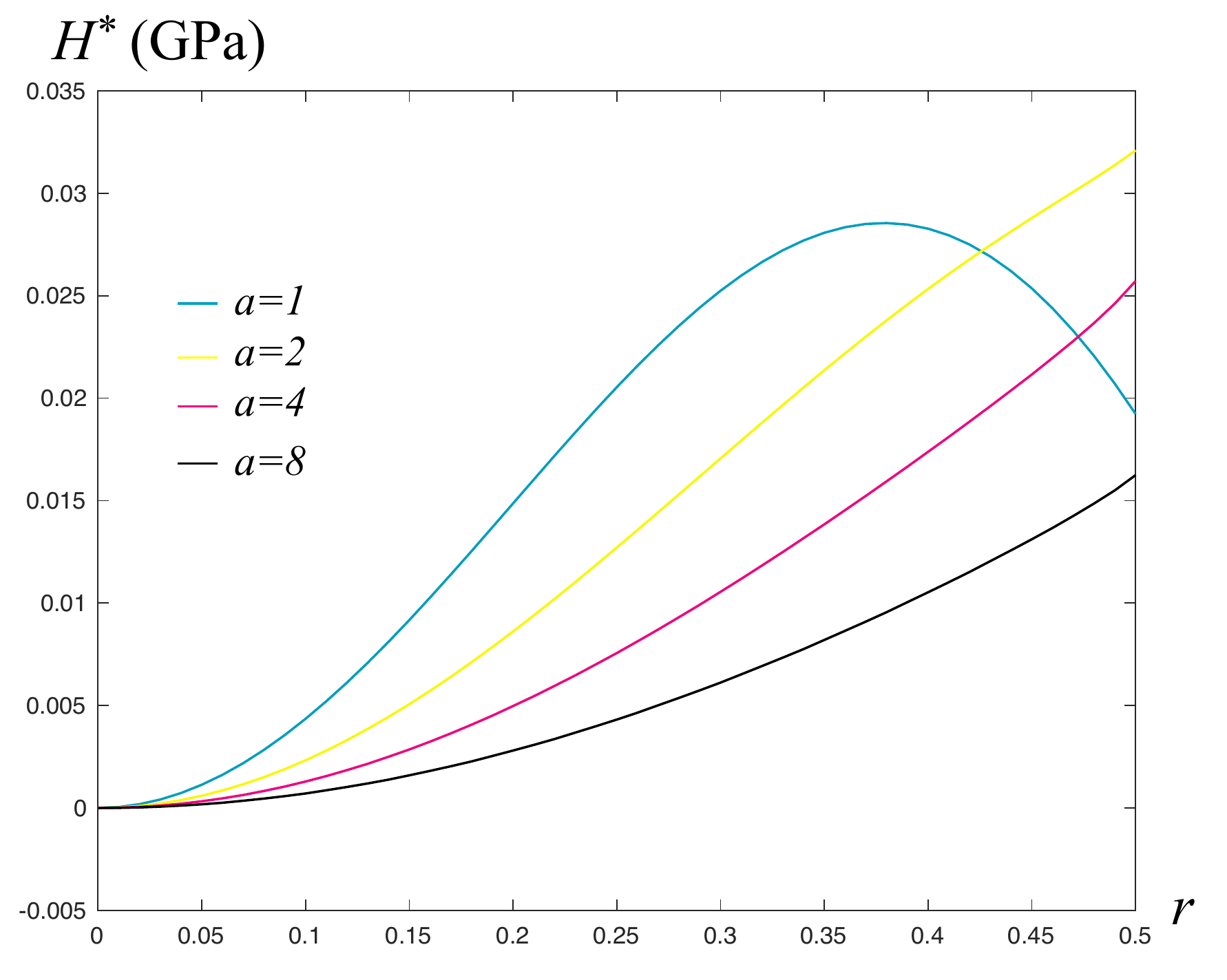}
	\caption{The plot of $H^*$ versus the largest half-axis of fibers of elliptical cross sections, where $E_1=1\,$GPa, $E_2=2\,$GPa, $\nu_1=0.1$, $\nu_2=0.4$: (i) $a=1$ (blue), (ii) $a=2$ (orange), (iii) $a=4$ (red), (iv) $a=8$ (black).}
	\label{fig:2}
\end{figure}

For the FRC bar it will be shown in the next Section that the minimum of functional \eqref{eq:comp17} is $\frac{1}{2}(F^*-\nu^*_{\alpha \beta}C^*_{\alpha \beta \gamma \delta}\nu^*_{\gamma \delta}) h^2\langle \xi ^2\rangle $. Therefore it makes sense to investigate the quantity $H^*=F^*-\nu^*_{\alpha \beta}C^*_{\alpha \beta \gamma \delta}\nu^*_{\gamma \delta}$ giving the correction to the stiffnesses on extension and bending. We take the microstructure of the composite in such a way that the cross sections of the fibers are ellipses of half-axes $r$ and $r/a$ placed in the middle of the unit quadratic periodic cells. Then the region occupied by one fiber in the unit cell is given by the equation
\begin{displaymath}
\frac{(z_1-1/2)^2}{r^2}+\frac{(z_2-1/2)^2}{(r/a)^2}\le 1.
\end{displaymath}
Note that for the circle with $a=1$ the effective Poisson ratio tensor is $\nu^*_{\alpha \beta}=\nu^* \delta_{\alpha \beta}$, however this property is no longer valid for ellipses with $a>1$. Assigning the index 1 to the matrix and index 2 to the fiber, we choose the Young moduli and Poisson ratios as follows: $E_1=1\,$GPa, $E_2=2\,$GPa, $\nu_1=0.1$, $\nu_2=0.4$. The plot of $H^*$ as function of the largest half-axis $r$ for four different aspect ratios of the ellipses $a=1,2,4,8$ is presented in Fig.~\ref{fig:2}. We see that the correction $H^*$ varies in the range $(0,0.03)\,$GPa which is not so large compared with the mean value of the Young moduli. Since the volume fraction of the fibers is $s_2=\pi r^2/a$, we can also plot $H^*$ as function of this volume fraction.

\begin{figure}[htb]
	\centering
	\includegraphics[width=7cm]{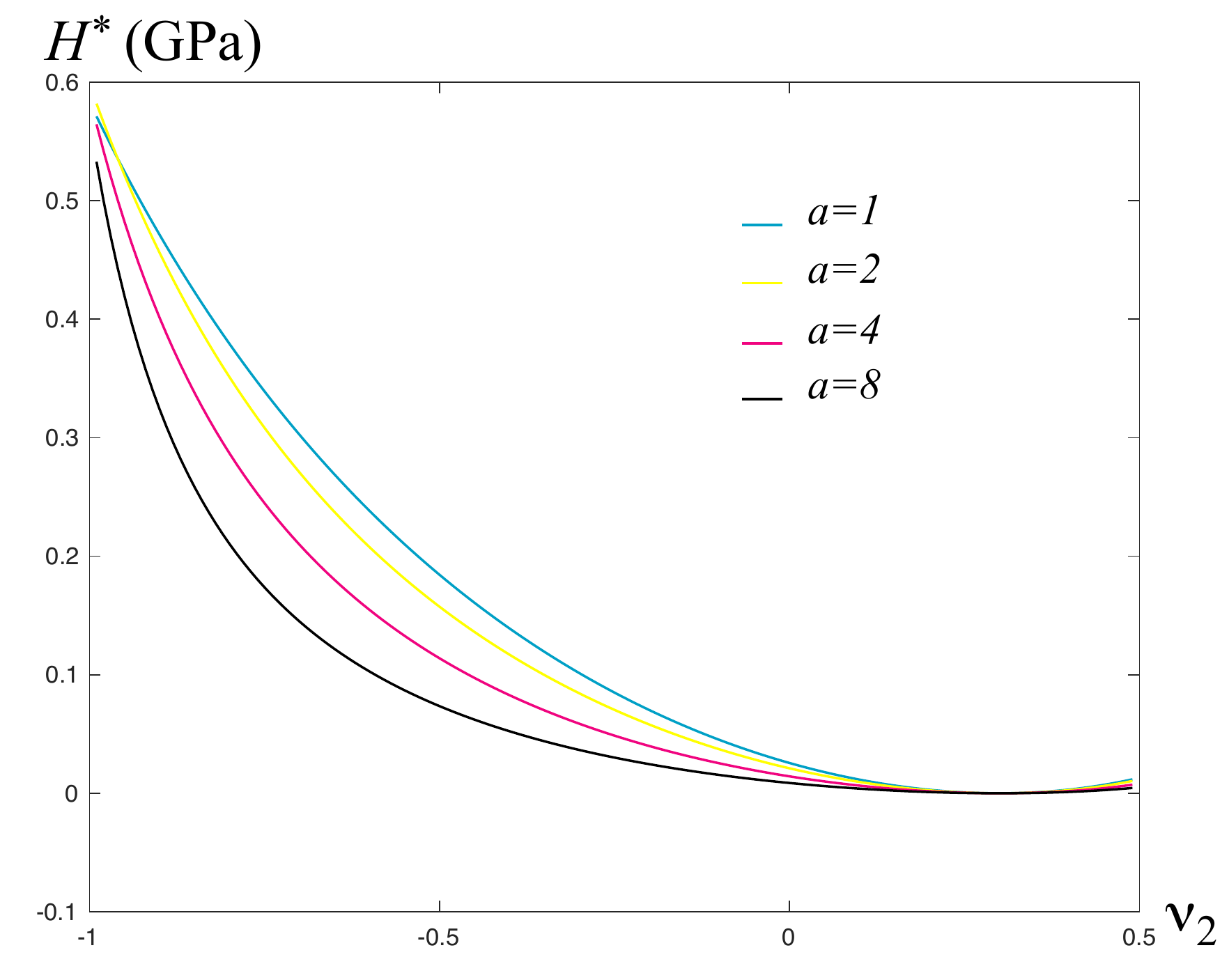}
	\caption{The plot of $H^*$ versus $\nu_2$ of fibers of elliptical cross sections, where $r=0.4$, $E_1=1\,$GPa, $E_2=2\,$GPa, $\nu_1=0.3$: (i) $a=1$ (blue), (ii) $a=2$ (orange), (iii) $a=4$ (red), (iv) $a=8$ (black).}
	\label{fig:3}
\end{figure}

Let us now fix the largest half-axis of elliptical cross section of the fibers to be $r=0.4$, choose $E_1=1\,$GPa, $E_2=2\,$GPa, $\nu_1=0.3$, vary the Poisson ratio of the fiber $\nu_2$ in its admissible range between $-1$ and $0.5$, and plot $H^*$ as function of $\nu_2$. Looking at this plot shown in Fig.~\ref{fig:3} we see that the larger the difference between the Poisson ratios, the larger is the correction $H^*$. For $\nu_2=0.3$ the correction $H^*$ vanishes as expected. For $\nu_2$ near $-1$ which is not unrealistic \citep{milton1992composite} the correction is more than one third of the mean Young modulus which can no longer be neglected.

\begin{figure}[htb]
	\centering
	\includegraphics[width=7cm]{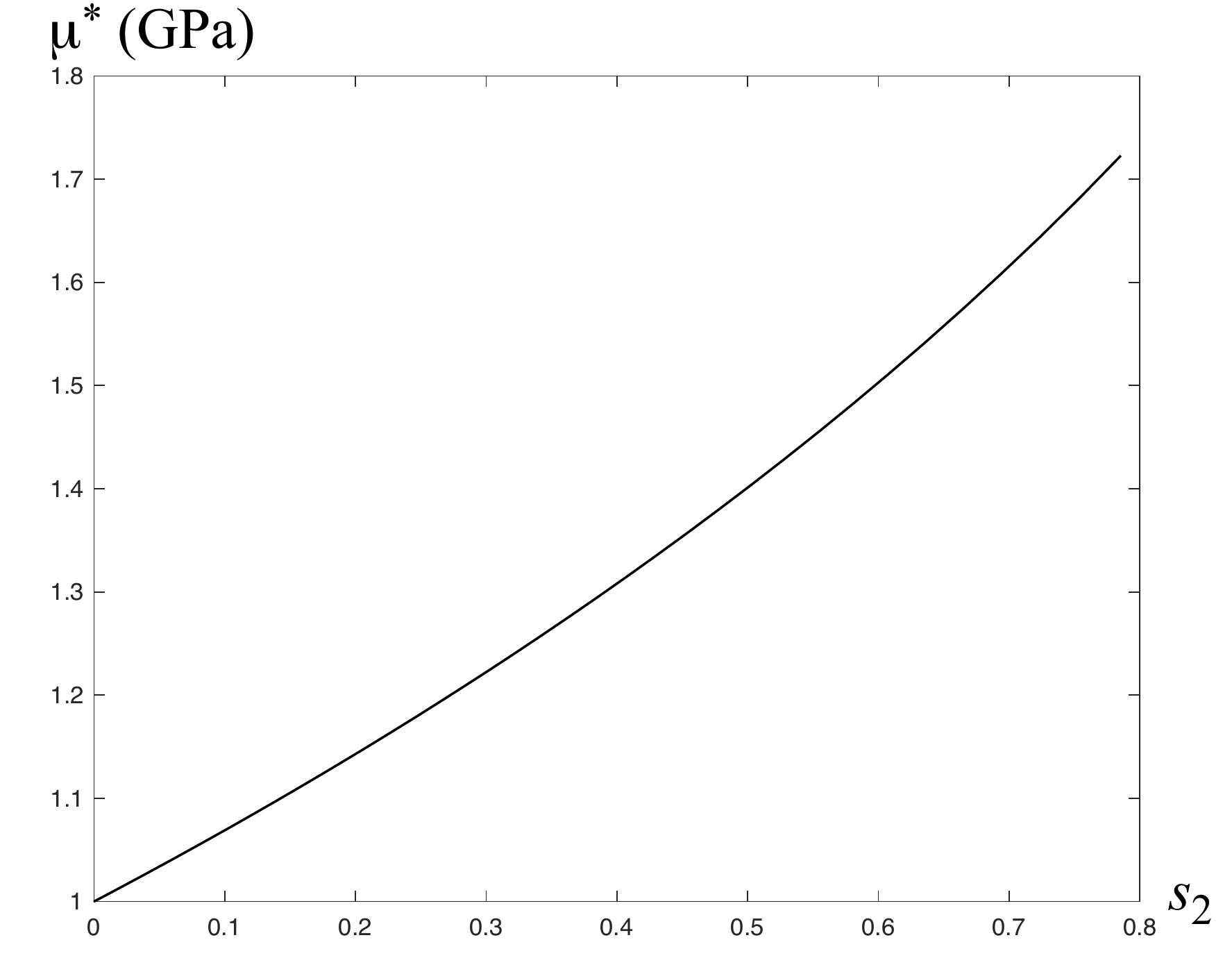}
	\caption{The plot of $\mu ^*$ versus the volume fraction $s_2=\pi r^2$ of fibers of circular cross section, where $\mu_1=1\,$GPa, $\mu_2=2\,$GPa.}
	\label{fig:4}
\end{figure}

The solution of the cell problem \eqref{eq:comp10} by the finite element method is quite similar. Its solution can be obtained from the previous problem if we put $\lambda (\vb{z})=0$. Below we present the numerical solution for the fibers with circular cross sections periodically embedded in the matrix. The unit cell is a square of length 1 with the circle of radius $r$ in the middle representing the cross section of the fiber. In this case the effective tensor is $\mu^*_{\alpha \beta }=\mu^*\delta _{\alpha \beta }$. The plot of $\mu^*$ as function of $s_2=\pi r^2$ is presented in Fig.~\ref{fig:4}. For the numerical simulation we choose $\mu_1=1\,$GPa, $\mu_2=2\,$GPa. 

\section{Solution of the homogenized cross-sectional problems}

\begin{figure}[htb]
	\centering
	\includegraphics[width=7cm]{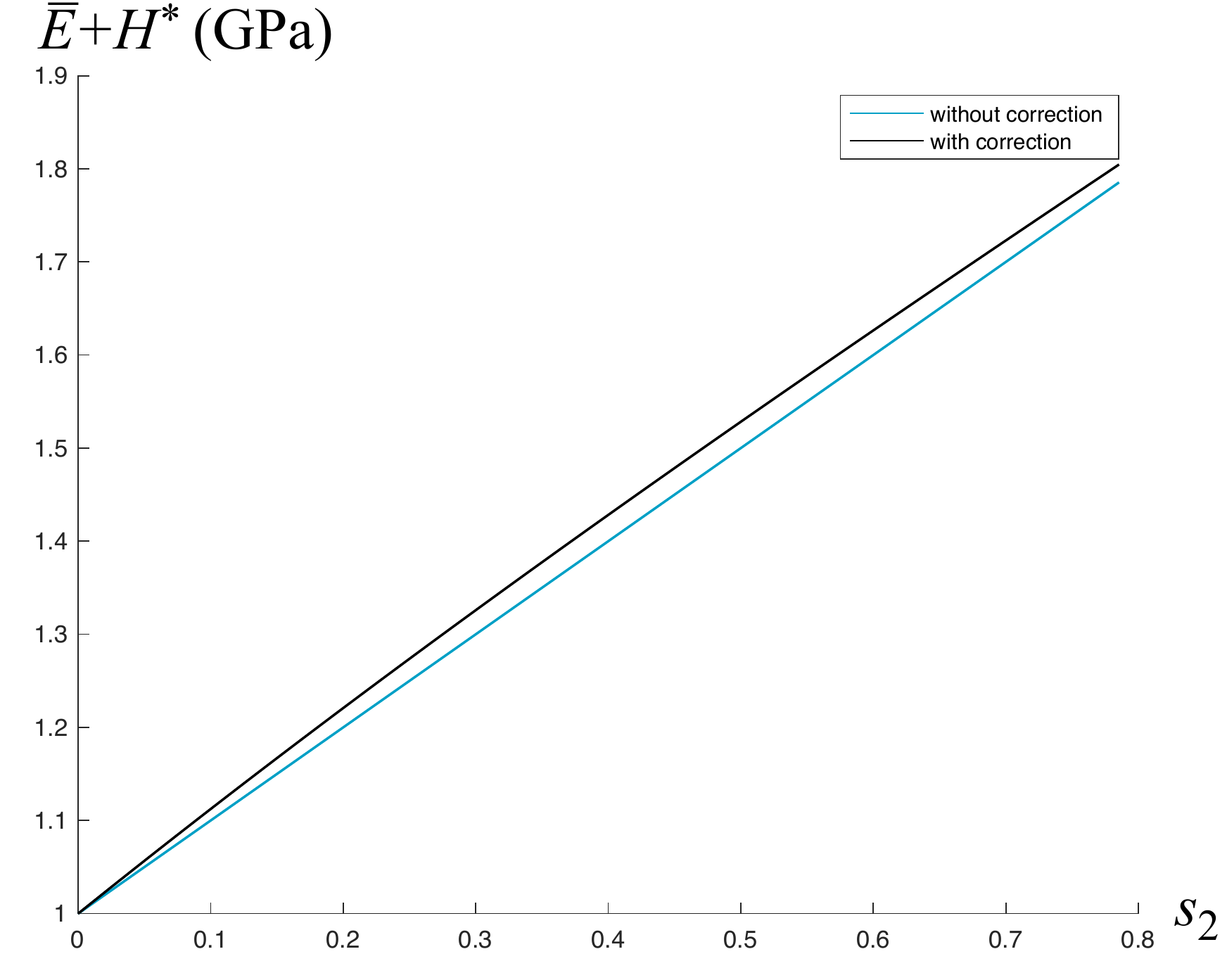}
	\caption{The plot of $\bar{E}$ and $\bar{E}+H^*$ versus the volume fraction $s_2=\pi r^2$ of fibers of circular cross section, where $E_1=1\,$GPa, $E_2=2\,$GPa, $\nu_1=0.1$, $\nu_2=0.4$.}
	\label{fig:5}
\end{figure}

Let us first consider the homogenized cross-sectional problem \eqref{eq:comp17} with the average transverse energy being given by \eqref{eq:comp21}. It is convenient to present the latter formula in the form
\begin{equation}
\label{eq:hcs1}
\bar{W}_\perp(\bar{\varepsilon}_{\alpha \beta},\xi)=\frac{1}{2}C^*_{\alpha \beta \gamma \delta}(\bar{\varepsilon}_{\alpha \beta}+\nu^*_{\alpha \beta}\xi)(\bar{\varepsilon}_{\gamma \delta}+\nu^*_{\gamma \delta}\xi)+\frac{1}{2}H^* \xi ^2,
\end{equation}
where $\xi =\gamma + h\Omega_\alpha y_\alpha $. Note that the second term in \eqref{eq:hcs1} does not depend on $\psi_\alpha (\vb{y})$, so we need to minimize functional \eqref{eq:comp17} that contains just the first term. We want to show that the minimum of the functional \eqref{eq:comp17} that contains just the first term among $\psi_\alpha (\vb{y})$ satisfying the constraints \eqref{eq:comp17a} is equal to zero. Indeed, following the same line of reasoning as in Section 4 we choose the minimizer in the form
\begin{equation*}
\check{\psi }_\alpha =-\nu ^*_{\alpha \beta}\gamma y_\beta -\frac{1}{2} a_{\alpha \beta \gamma}( y_\beta y_\gamma -\langle y_\beta y_\gamma\rangle /|\bar{\mathcal{A}}|) ,
\end{equation*}
where
\begin{displaymath}
a_{\alpha \beta \gamma}=h(\nu ^*_{\alpha \beta}\Omega_\gamma +\nu ^*_{\alpha \gamma}\Omega_\beta-\nu ^*_{\beta \gamma}\Omega_\alpha ).
\end{displaymath}
It is easy to check that $\check{\psi }_\alpha $ satisfy the constraints \eqref{eq:comp17a} and that the first term evaluated at these functions vanishes identically, so 
\begin{equation*}
\min_{\psi_\alpha (\vb{y})\in \eqref{eq:comp17a}} \bar{I}_\perp[\psi_\alpha (\vb{y})] = \frac{1}{2}H^* h^2\langle \xi ^2\rangle .
\end{equation*}
Adding the average longitudinal energy density $h^2\langle W_\parallel \rangle =\frac{1}{2} \bar{E}\langle \xi ^2\rangle $ to this average transverse energy density and integrating $\xi^2=(\gamma +h\Omega_\alpha y_\alpha)^2$ over the cross section, we find the final energy density of extension and bending of FRC beam in the form
\begin{equation*}
\Phi (\gamma ,\Omega _\alpha)=\frac{1}{2} (\bar{E}+H^*)(|\mathcal{A}|\gamma ^2+I_{\alpha \beta}\Omega _\alpha \Omega_\beta ),
\end{equation*}
where
\begin{equation*}
\bar{E}=\llangle E\rrangle =s_1E_1+s_2E_2,\quad I_{\alpha \beta }=h^4 \langle y_\alpha y_\beta \rangle .
\end{equation*}
Thus, $H^*$ is the correction to the stiffnesses on extension and bending of FRC beam. To see how this correction changes the stiffnesses on extension and bending of FRC beam we plot both $\bar{E}$ and $\bar{E}+H^*$ as function of the volume fraction of fibers with circular cross sections placed in the middle of the quadratic periodic cell (see Fig.~\ref{fig:5}). The chosen elastic moduli are: $E_1=1\,$GPa, $E_2=2\,$GPa, $\nu_1=0.1$, $\nu_2=0.4$. The correction is about 3 percent of $\bar{E}$ in this case. Note that this correction becomes one third of $\bar{E}$ if $\nu_2$ is near $-1$.  

We now turn to the minimization problem \eqref{eq:comp11}. First of all it is easy to see that the constraint $\langle \psi \rangle =0$ does not affect the minimum value of \eqref{eq:comp11}, because the latter is invariant with respect to the change of unknown function $\psi \to \psi +c$, with $c$ being a constant. By such the change one can always achieve the fulfillment of the constraint $\langle \psi \rangle =0$. The minimizer $\check{\psi }$ of $\bar{I}_\angle$ is proportional to $h\Omega $. Therefore the torsional rigidity $C$ can be calculated by
\begin{equation}
\label{eq:hcs5}
C=h^4\inf_{\bar{\psi }} \, \langle \mu^* \delta _{\alpha \beta }
(\bar{\psi }_{|\alpha }-e_{\alpha \sigma }y_\sigma )(\bar{\psi }_{|\beta}-e_{\beta \kappa }y_\kappa ) \rangle ,
\end{equation}
where $\bar{\psi}=\psi /h\Omega$. The solution to this problem is well-known for the elliptical and rectangular cross sections \citep{le1999vibrations}. For the FRC beam with the elliptical cross section described by the equation 
\begin{eqnarray*}
c_{\alpha \beta }y_\alpha y_\beta \le 1,
\end{eqnarray*}
where $c_{\alpha \beta }$ are the components of a positive definite symmetric second-rank tensor, the torsional rigidity  reads 
\begin{equation}
C=\frac{4\mu ^*}{c^2}c_{\mu \lambda }c_{\lambda \tau} I_{\mu \tau },
\label{eq:hcs6}
\end{equation}
where $I_{\mu \tau }=h^4\langle y_\mu y_\tau \rangle $ are the moments of inertia of the cross section.

In the co-ordinate system associated with the principal axes of the ellipse
\begin{gather*}
c_{11}=\frac{h^2}{b_1^2}, \quad c_{22}=\frac{h^2}{b_2^2},
\quad c_{12}=0, \\
I_{11}=\frac{1}{4}\pi b_1^3b_2, \quad 
I_{22}=\frac{1}{4}\pi b_2^3b_1, \quad I_{12}=0,
\end{gather*}
where $b_1,b_2$ are the half-lengths of the major and minor axes. Substituting these formulas into \eqref{eq:hcs6} we get finally
\begin{eqnarray*}
C=\frac{\pi \mu^* b_1^3 b_2^3}{b_1^2+b_2^2}=\frac{4\mu^* }{(I^{-1})_{
\alpha \alpha} },
\end{eqnarray*}
where $(I^{-1})_{\alpha \beta }$ is a tensor inverse to $I_{\alpha \beta }$.

It is easy to see that \eqref{eq:hcs6} is also the solution to the problem \eqref{eq:hcs5} for a hollow elliptical cross section
\[
\lambda ^2\le c_{\alpha \beta }y_\alpha y_\beta \le 1,
\quad \lambda <1.
\]
Similar calculations give
\begin{eqnarray*}
C=\frac{\pi \mu^* (1-\lambda ^4) b_1^3 b_2^3}{b_1^2+b_2^2}.
\end{eqnarray*}

For the rectangular cross section of width $a$ and height 1, where $a<1$, the solution is found in form of an infinite series \citep{le1999vibrations}. The torsional rigidity is given by the formula
\begin{eqnarray*}
C=\mu ^* c h^4a^3 ,
\end{eqnarray*}
where
\begin{eqnarray*}
c=\frac{1}{3}-\frac{64a}{\pi ^5}\sum_{n=0}^\infty
\frac{\tanh (\pi (2n+1)/2a)}{(2n+1)^5}.
\end{eqnarray*}
For other cross sections the problem \eqref{eq:hcs5} can be solved by the finite element method.

\section{1-D beam theory}

Summarizing the results obtained in Sections 3-6, we can now reduce the 3-D problem of equilibrium of the FRC beam to the following 1-D variational problem: Minimize the energy functional
\begin{equation*}
J[\vb{v},\varphi ]=\int_0^L \Phi(\gamma ,\Omega_\alpha,\Omega)\dd{x}-\vb{f}\vdot \vb{v}(L)+m_\alpha v_{\alpha ,x}(L)-m \varphi (L)
\end{equation*}
among functions $\vb{v}$ and $\varphi$ satisfying the kinematic boundary conditions
\begin{equation*}
\vb{v}(0)=0, \quad v_{\alpha ,x}(0)=0,\quad \varphi(0)=0.
\end{equation*}
In this functional the stored energy density $\Phi(\gamma,\Omega_\alpha,\Omega)$ is given by
\begin{equation*}
\Phi(\gamma,\Omega_\alpha,\Omega)=\frac{1}{2}(\bar{E}+H^*)(|\mathcal{A}|\gamma ^2+I_{\alpha \beta}\Omega _\alpha \Omega_\beta )+\frac{1}{2}C\Omega^2,
\end{equation*}
while the resultant forces and moments acting at $x=L$ are equal to
\begin{align*}
\vb{f}&=\int_{\mathcal{A}}\vb{t}(x_1,x_2)\dd[2]{x},
\\
m_\alpha&=\int_{\mathcal{A}} x_\alpha t(x_1,x_2)\dd[2]{x},\quad m=\int_{\mathcal{A}} e_{\alpha \beta }x_\alpha t_\beta (x_1,x_2)\dd[2]{x}.
\end{align*}

The standard calculus of variations shows that $\vb{v}$ and $\varphi$ satisfy the equilibrium equations
\begin{equation}
\label{eq:1D4}
T_{,x}=0, \quad M_{\alpha ,xx}=0,\quad M_{,x}=0,
\end{equation}
where
\begin{equation*}
T=\frac{\partial \Phi}{\partial \gamma}=(\bar{E}+H^*)|\mathcal{A}|\gamma,\quad M_\alpha=\frac{\partial \Phi}{\partial \Omega_\alpha }= (\bar{E}+H^*)I_{\alpha \beta}\Omega_\beta,\quad M=\frac{\partial \Phi}{\partial \Omega}=C\Omega.
\end{equation*}
Besides, the following boundary conditions must be fulfilled at $x=L$:
\begin{equation}
\label{eq:1D6}
T(L)=f,\quad M_\alpha(L)=m_\alpha ,\quad M_{\alpha,x}=f_\alpha ,\quad M=m.
\end{equation}
Using the technique of Gamma-convergence \citep{braides2002gamma,milton2003theory}, one can prove that the solution of \eqref{eq:1D4}-\eqref{eq:1D6} converges to the minimizer of \eqref{2.2} in the energetic norm as $h/L\to 0$ and $\epsilon=c/h\to 0$.

It is easy to extend this one-dimensional theory to the dynamics of FRC beam, where functions $\vb{v}$ and $\varphi$ depend on $x$ and $t$. One need just to include into the one-dimensional functional the kinetic energy density which, after the dimension reduction and homogenization in accordance with \eqref{5.2}, takes the form
\begin{equation}
\label{eq:1D7}
\Theta =\frac{1}{2}\bar{\rho}(|\mathcal{A}|\dot{\vb{v}}\vdot \dot{\vb{v}}+I_{\alpha \alpha
}\dot{\varphi}^2), 
\end{equation}
with $\bar{\rho}=\llangle \rho \rrangle =s_1\rho_1+s_2\rho_2$ being the mass density averaged over the unit cell. Hamilton's variational principle for 1-D beam theory states that, among all admissible functions $\vb{v}(x,t)$ and $\varphi(x,t)$ satisfying the initial and end conditions as well as the kinematic boundary conditions, the true displacement $\check{\vb{v}}$ and rotation $\check{\varphi}$ are the extremal of the action functional
\begin{equation*}
J[\vb{v},\varphi ]=\int_{t_0}^{t_1}\int_0^L (\Theta -\Phi)\dd{x}\dd{t}+\int_{t_0}^{t_1}(\vb{f}\vdot \vb{v}(L)-m_\alpha v_{\alpha ,x}(L)+m \varphi (L))\dd{t}.
\end{equation*}
The Euler equations become
\begin{equation*}
\bar{\rho}|\mathcal{A}|\ddot{v}=T_{,x}, \quad \bar{\rho}|\mathcal{A}|\ddot{v}_\alpha=M_{\alpha ,xx},\quad \bar{\rho}I_{\alpha \alpha
}\ddot{\varphi}=M_{,x}.
\end{equation*}
These equations are subjected to the boundary conditions \eqref{eq:1D7} and the initial conditions. Similar to the static case, it can be proved that the solution of this 1-D dynamic theory converges to the solution of the 3-D theory in the limits $h/L\to 0$ and $\epsilon=c/h\to 0$ and for low frequency vibrations \citep{le1999vibrations}.

\section{Conclusion}
It is shown in this paper that the rigorous first order approximate 1-D theory of thin FRC beams can be derived from the exact 3-D elasticity theory by the variational-asymptotic method. The developed finite element code can be used to solve cross-sectional problems with arbitrary elastic moduli of the fibers and matrix as well as arbitrary distributions and shapes of fibers. The extension of this multi-scaled asymptotic analysis to curved and naturally twisted FRC beams is straightforward \citep{le1999vibrations}. As seen from \citep{le1999vibrations}, the extension, bending, and torsion modes of beams are coupled in that case. The VAM combined with the finite element cross-sectional analysis can also be applied to the nonlinear FRC beams with periodic microstructure in the spirit of \citep{yu2002on,yu2005a,yu2007variational,liu2017two}. Finally, let us mention the FRC beams with randomly distributed fibers under torsion analyzed in \citep{barretta2015on}. The analysis of extension and bending of such beams requires the solution of the plane strain problem which is quite challenging. 

\section*{Appendix: Matlab-code}
\begin{lstlisting}
function [CH,DH,FH] = homogenizecs(lx, ly, lambda, mu, phi, x)
%%%%%%%%%%%%%%%%%%%%%%%%%%%%%%%%%%%%%%%%%%%%%%%%%%%%%%%%%%%%%%%%%%%%%%%%
% lx = Unit cell length in x-direction.
% ly = Unit cell length in y-direction.
% lambda = Lame's first parameter for both materials. Two entries.
% mu = Lame's second parameter for both materials. Two entries.
% phi = Angle between horizontal and vertical cell wall. Degrees
% x = Material indicator matrix. Size used to determine nelx/nely
%%%%%%%%%%%%%%%%%%%%%%%%%%%%%%%%%%%%%%%%%%%%%%%%%%%%%%%%%%%%%%%%%%%%%%%%
%% INITIALIZE
% Compute contraction ratios
nu = [lambda(1)/(2*(lambda(1) + mu(1)))...
    lambda(2)/(2*(lambda(2) + mu(2)))];
% Deduce discretization
[nely, nelx] = size(x);
% Stiffness matrix consists of two parts, one belonging to lambda and
% one belonging to mu. Same goes for load vector
dx = lx/nelx; dy = ly/nely;
nel = nelx*nely;
[keLambda, keMu, feLambda, feMu, feXiLambda, feXiMu] = elementMatVec(dx/2, dy/2, phi);
% Node numbers and element degrees of freedom for full (not periodic) mesh
nodenrs = reshape(1:(1+nelx)*(1+nely),1+nely,1+nelx);
edofVec = reshape(2*nodenrs(1:end-1,1:end-1)+1,nel,1);
edofMat = repmat(edofVec,1,8)+repmat([0 1 2*nely+[2 3 0 1] -2 -1],nel,1);
%% IMPOSE PERIODIC BOUNDARY CONDITIONS
% Use original edofMat to index into list with the periodic dofs
nn = (nelx+1)*(nely+1); % Total number of nodes
nnP = (nelx)*(nely); % Total number of unique nodes
nnPArray = reshape(1:nnP, nely, nelx);
% Extend with a mirror of the top border
nnPArray(end+1,:) = nnPArray(1,:);
% Extend with a mirror of the left border
nnPArray(:,end+1) = nnPArray(:,1);
% Make a vector into which we can index using edofMat:
dofVector = zeros(2*nn, 1);
dofVector(1:2:end) = 2*nnPArray(:)-1;
dofVector(2:2:end) = 2*nnPArray(:);
edofMat = dofVector(edofMat);
ndof = 2*nnP; % Number of dofs
%% ASSEMBLE STIFFNESS MATRIX
% Indexing vectors
iK = kron(edofMat,ones(8,1))';
jK = kron(edofMat,ones(1,8))';
% Material properties in the different elements
lambda = lambda(1)*(x==1) + lambda(2)*(x==2);
mu = mu(1)*(x==1) + mu(2)*(x==2);
nu = nu(1)*(x==1) + nu(2)*(x==2);
% The corresponding stiffness matrix entries
sK = keLambda(:)*lambda(:).' + keMu(:)*mu(:).';
K = sparse(iK(:), jK(:), sK(:), ndof, ndof);
%% LOAD VECTORS AND SOLUTION
% Assembly three load cases corresponding to the three strain cases
sF = feLambda(:)*lambda(:).'+feMu(:)*mu(:).';
lambda_nu = lambda.*nu;
mu_nu = mu.*nu;
sF=[sF; feXiLambda(:)*lambda_nu(:).' + feXiMu(:)*mu_nu(:).'];
iF = repmat(edofMat',4,1);
jF = [ones(8,nel); 2*ones(8,nel); 3*ones(8,nel); 4*ones(8,nel)];
F = sparse(iF(:), jF(:), sF(:), ndof, 4);
% Solve (remember to constrain one node)
chi(3:ndof,:) = K(3:ndof,3:ndof)\F(3:ndof,:);
chi_4=chi(:,4);
%% HOMOGENIZATION
% The displacement vectors corresponding to the unit strain cases
chi0 = zeros(nel, 8, 3);
% The element displacements for the three unit strains
chi0_e = zeros(8, 4);
ke = keMu + keLambda; % Here the exact ratio does not matter, because
fe = feMu + feLambda; % it is reflected in the load vector
fe = [fe feXiLambda+feXiMu];
chi0_e([3 5:end],:) = ke([3 5:end],[3 5:end])\fe([3 5:end],:);
chi0_4_e=nu(:)*chi0_e(:,4)';
% epsilon0_11 = (1, 0, 0)
chi0(:,:,1) = kron(chi0_e(:,1)', ones(nel,1));
% epsilon0_22 = (0, 1, 0)
chi0(:,:,2) = kron(chi0_e(:,2)', ones(nel,1));
% epsilon0_12 = (0, 0, 1)
chi0(:,:,3) = kron(chi0_e(:,3)', ones(nel,1));
CH = zeros(3);
DH = [0; 0; 0];
cellVolume = lx*ly;
sumLambda = ((chi0_4_e(:,:)-chi_4(edofMat))*...
    keLambda).*(chi0_4_e(:,:)-chi_4(edofMat));
sumMu = ((chi0_4_e(:,:)-chi_4(edofMat))*keMu).*...
    (chi0_4_e(:,:)-chi_4(edofMat));
sumLambda = reshape(sum(sumLambda,2), nely, nelx);
sumMu = reshape(sum(sumMu,2), nely, nelx);
FH = 1/cellVolume*sum(sum(lambda.*sumLambda+mu.*sumMu));
for i = 1:3
    sumLambda = ((chi0_4_e(:,:)-chi_4(edofMat))*...
    keLambda).*(chi0(:,:,i)-chi(edofMat+(i-1)*ndof));
    sumMu = ((chi0_4_e(:,:)-chi_4(edofMat))*keMu).*...
    (chi0(:,:,i)-chi(edofMat+(i-1)*ndof));
    sumLambda = reshape(sum(sumLambda,2), nely, nelx);
    sumMu = reshape(sum(sumMu,2), nely, nelx);
    DH(i) = 1/cellVolume*sum(sum(lambda.*sumLambda+mu.*sumMu));
    for j = 1:3
        sumLambda = ((chi0(:,:,i) - chi(edofMat+(i-1)*ndof))*keLambda).*...
        (chi0(:,:,j) - chi(edofMat+(j-1)*ndof));
        sumMu = ((chi0(:,:,i) - chi(edofMat+(i-1)*ndof))*keMu).*...
        (chi0(:,:,j) - chi(edofMat+(j-1)*ndof));
        sumLambda = reshape(sum(sumLambda,2), nely, nelx);
        sumMu = reshape(sum(sumMu,2), nely, nelx);
        % Homogenized elasticity tensor
        CH(i,j) = 1/cellVolume*sum(sum(lambda.*sumLambda + mu.*sumMu));
    end
end
disp('--- Homogenized elasticity CH ---'); disp(CH)
disp('--- Homogenized elasticity DH ---'); disp(DH)
disp('--- Homogenized elasticity FH ---'); disp(FH)
%% COMPUTE ELEMENT STIFFNESS MATRIX AND FORCE VECTOR (NUMERICALLY)
function [keLambda, keMu, feLambda, feMu, feXiLambda, feXiMu] = elementMatVec(a, b, phi)
% Constitutive matrix contributions
CMu = diag([2 2 1]); CLambda = zeros(3); CLambda(1:2,1:2) = 1; 
% Two Gauss points in both directions
xx=[-1/sqrt(3), 1/sqrt(3)]; yy = xx;
ww=[1,1];
% Initialize
keLambda = zeros(8,8); keMu = zeros(8,8);
feLambda = zeros(8,3); feMu = zeros(8,3); feXiLambda = zeros(8,1);feXiMu = zeros(8,1);
L = zeros(3,4); L(1,1) = 1; L(2,4) = 1; L(3,2:3) = 1;
for ii=1:length(xx)
for jj=1:length(yy)
% Integration point
x = xx(ii); y = yy(jj);
% Differentiated shape functions
dNx = 1/4*[-(1-y) (1-y) (1+y) -(1+y)];
dNy = 1/4*[-(1-x) -(1+x) (1+x) (1-x)];
% Jacobian
J = [dNx; dNy]*[-a a a+2*b/tan(phi*pi/180) 2*b/tan(phi*pi/180)-a; ...
-b -b b b]';
detJ = J(1,1)*J(2,2) - J(1,2)*J(2,1);
invJ = 1/detJ*[J(2,2) -J(1,2); -J(2,1) J(1,1)];
% Weight factor at this point
weight = ww(ii)*ww(jj)*detJ;
% Strain-displacement matrix
G = [invJ zeros(2); zeros(2) invJ];
dN = zeros(4,8);
dN(1,1:2:8) = dNx;
dN(2,1:2:8) = dNy;
dN(3,2:2:8) = dNx;
dN(4,2:2:8) = dNy;
B = L*G*dN;
% Element matrices
keLambda = keLambda + weight*(B' * CLambda * B);
keMu = keMu + weight*(B' * CMu * B);
% Element loads
feLambda = feLambda + weight*(B' * CLambda * diag([1 1 1]));
feMu = feMu + weight*(B' * CMu * diag([1 1 1]));
feXiLambda = feXiLambda + weight*(B' * CLambda* [1; 1; 0]);
feXiMu=feXiMu+ weight*(B' * CMu* [1; 1; 0]);
end
end
\end{lstlisting}


\begin{thebibliography}{56}
\expandafter\ifx\csname natexlab\endcsname\relax\def\natexlab#1{#1}\fi
\expandafter\ifx\csname url\endcsname\relax
  \def\url#1{\texttt{#1}}\fi
\expandafter\ifx\csname urlprefix\endcsname\relax\def\urlprefix{URL }\fi

\bibitem[{Andreasen(2014)}]{andreasen2014how}
Andreassen, E., Andreasen, C.~S., 2014. How to determine composite material properties using numerical homogenization. Computational Materials Science 83, 488--495.

\bibitem[{Barretta et al.(2015)}]{barretta2015on}
Barretta, R., Luciano, R., Willis, J.~R., 2015. On torsion of random composite beams. Composite Structures 132, 915--922.

\bibitem[{Berdichevsky(1979)}]{berdichevsky1979variational}
Berdichevsky, V.~L., 1979. Variational-asymptotic method of constructing a theory of shells. Journal of Applied Mathematics and Mechanics 43~(4), 711--736.

\bibitem[{Berdichevsky(1981)}]{berdichevsky1981on}  
Berdichevsky, V.~L., 1981. On the energy of an elastic rod. Journal of Applied Mathematics and Mechanics 45(4), 518--529.

\bibitem[{Berdichevsky(2009)}]{berdichevsky2009variational}
Berdichevsky, V.~L., 2009. {Variational Principles of Continuum Mechanics, vols. 1 and 2}. Springer Verlag, Berlin.

\bibitem[{Berdichevsky(2010{\natexlab{a}})}]{berdichevsky2010asymptotic}
Berdichevsky, V.~L., 2010{\natexlab{a}}. An asymptotic theory of sandwich plates. International Journal of Engineering Science 48~(3), 383--404.

\bibitem[{Berdichevsky(2010{\natexlab{b}})}]{berdichevsky2010nonlinear}
Berdichevsky, V.~L., 2010{\natexlab{b}}. Nonlinear theory of hard-skin plates and shells. International Journal of Engineering Science 48~(3), 357--369.

\bibitem[{Braides(2002)}]{braides2002gamma}
Braides, A., 2002. Gamma-convergence for Beginners. Vol.~22. Clarendon Press.

\bibitem[Chandra et al.(1999)]{chandra1999damping}
Chandra, R., Singh, S.~P., Gupta, K., 1999. Damping studies in fiber-reinforced composites--a review. Composite Structures 46(1), 41--51.

\bibitem[Hodges(1990)]{hodges1990review}
Hodges, D.~H., 1990. Review of composite rotor blade modeling. AIAA journal 28(3), 561--565.

\bibitem[Hodges et al.(1992)]{hodges1992on}
Hodges, D.~H., Atilgan, A.~R., Cesnik, C.~E., Fulton, M.~V., 1992. On a simplified strain energy function for geometrically nonlinear behaviour of anisotropic beams. Composites Engineering 2(5-7), 513--526.

\bibitem[Hughes(2012)]{hughes2012the}
Hughes, T.~J., 2012. The finite element method: linear static and dynamic finite element analysis. Courier Corporation.

\bibitem[{Le(1986{\natexlab{b}})}]{Le86a}
Le, K.~C., 1986{\natexlab{b}}. The theory of piezoelectric shells. Journal of
  Applied Mathematics and Mechanics 50~(1), 98--105.

\bibitem[{Le(1999)}]{le1999vibrations}
Le, K.~C., 1999. Vibrations of shells and rods. Springer Verlag.

\bibitem[{Le and Yi(2016)}]{le2016asymptotically}
Le, K.~C., Yi, J.-H., 2016. An asymptotically exact theory of smart sandwich
  shells. International Journal of Engineering Science 106, 179--198.

\bibitem[{Le(2017)}]{le2017an}
Le, K.~C., 2017. An asymptotically exact theory of functionally graded piezoelectric shells. International Journal of Engineering Science 112, 42-62.

\bibitem[{Lee and Hodges(2009{\natexlab{a}})}]{lee2009adynamic}
Lee, C.-Y., Hodges, D.~H., 2009{\natexlab{a}}. {Dynamic variational-asymptotic
  procedure for laminated composite shells - Part I: Low-frequency vibration
  analysis}. Journal of Applied Mechanics 76~(1), 011002.

\bibitem[{Lee and Hodges(2009{\natexlab{b}})}]{lee2009bdynamic}
Lee, C.-Y., Hodges, D.~H., 2009{\natexlab{b}}. {Dynamic variational-asymptotic
  procedure for laminated composite shells - Part II: High-frequency vibration
  analysis}. Journal of Applied Mechanics 76~(1), 011003.

\bibitem[Librescu and Song(2005)]{librescu2005thin}  
Librescu, L. and Song, O., 2005. Thin-walled composite beams: theory and application (Vol. 131). Springer Science \& Business Media.

\bibitem[Liu et al.(2017)]{liu2017two} 
Liu, X., Rouf, K., Peng, B. and Yu, W., 2017. Two-step homogenization of textile composites using mechanics of structure genome. Composite Structures 171, 252--262.

\bibitem[Milton(1992)]{milton1992composite}
Milton, G.~W., 1992. Composite materials with Poisson's ratios close to -1. Journal of the Mechanics and Physics of Solids 40(5), 1105--1137.

\bibitem[{Milton(2003)}]{milton2003theory}  
Milton, G.W., 2003. Theory of Composites. Cambridge Monographs on Applied and Computational Mathematics.

\bibitem[Muskhelishvili(2013)]{muskhelishvili2013some} 
Muskhelishvili, N.~I., 2013. Some basic problems of the mathematical theory of elasticity. Springer Science \& Business Media.

\bibitem[{Rodriguez et al.(2001)}]{rodriguez2001closed}
Rodriguez-Ramos, R., Sabina, F.J., Guinovart-Diaz, R., Bravo-Castillero, J., 2001. Closed-form expressions for the effective coefficients of a fiber-reinforced composite with transversely isotropic constituents--I. Elastic and square symmetry. Mechanics of Materials 33(4), 223--235.

\bibitem[{Roy et~al.(2007)Roy, Yu, and Han}]{roy2007asymptotically}
Roy, S., Yu, W., Han, D., 2007. An asymptotically correct classical model for smart beams. International Journal of Solids and Structures 44~(25), 8424--8439.

\bibitem[{Sendeckyj(2016)}]{sendeckyj2016mechanics}
Sendeckyj, G.~P. (ed.), 2016. Mechanics of Composite Materials: Composite Materials (Vol. 2). Elsevier.

\bibitem[{Sigmund and Torquato(1997)}]{Sigmund1997}
Sigmund, O., Torquato, S., 1997. Design of materials with extreme thermal expansion using a three-phase topology optimization method. Journal of the Mechanics and Physics of Solids 45~(6), 1037--1067.

\bibitem[{Yu et al.(2002)}]{yu2002on}
Yu, W., Hodges, D.~H., Volovoi, V., Cesnik, C.~E., 2002. On Timoshenko-like modeling of initially curved and twisted composite beams. International Journal of Solids and Structures 39(19), 5101--5121.

\bibitem[{Yu et al.(2005)}]{yu2005a}
Yu, W., Hodges, D.~H., Volovoi, V.~V., Fuchs, E.~D., 2005. A generalized Vlasov theory for composite beams. Thin-Walled Structures 43(9), 1493--1511.

\bibitem[{Yu and Blair(2012)}]{yu2012a}
Yu, W. and Blair, M., 2012. GEBT: A general-purpose nonlinear analysis tool for composite beams. Composite Structures 94(9), 2677--2689.

\bibitem[{Yu and Tang(2007)}]{yu2007variational}
Yu, W., Tang, T., 2007. Variational asymptotic method for unit cell homogenization of periodically heterogeneous materials. International Journal of Solids and Structures 44(11-12), 3738--3755.

\end{thebibliography}
\end{document}